\documentclass[aps,amsmath,amssymb,nofootinbib]{revtex4} 
\usepackage{graphicx} 
\usepackage{amsmath}
\usepackage{amsfonts,amsbsy}
\usepackage{amssymb}

\def\gsim{ \,\, \vcenter{\hbox{$\buildrel{\displaystyle >}\over\sim$}}
 \,\,}
\def\lsim{ \,\, \vcenter{\hbox{$\buildrel{\displaystyle <}\over\sim$}}
 \,\,}
\def\be{\begin{equation}}
\def\ee{\end{equation}}
\def\bea{\begin{eqnarray}}
\def\eea{\end{eqnarray}}
\newcommand{\ud}{\, \mathrm{d}}
\newcommand{\xt}{{\mathbf{x}}}
\newcommand{\yt}{{\mathbf{y}}}
\newcommand{\rt}{{\mathbf{r}}}
\newcommand{\bt}{{\mathbf{b}}}
\newcommand{\ut}{{\mathbf{u}}}
\newcommand{\vt}{{\mathbf{v}}}
\newcommand{\kt}{{\mathbf{k}}}
\newcommand{\nabt}{\boldsymbol{\nabla}}
\newcommand{\tr}{\, \mathrm{tr} \, }

\begin{document}

\title{\bf Anisotropy of the semi-classical gluon field of a large
  nucleus at high energy}

\author{Adrian Dumitru}
\email{Adrian.Dumitru@baruch.cuny.edu}
\affiliation{Department of Natural Sciences, Baruch College, CUNY,
17 Lexington Avenue, New York, NY 10010, USA}
\affiliation{The Graduate School and University Center, The City University of New York, 365 Fifth Avenue, New York, NY 10016, USA}

\author{Vladimir Skokov}
\email{Vladimir.Skokov@wmich.edu}
\affiliation{Department of Physics, Western Michigan University, Kalamazoo, MI 49008, USA}

\begin{abstract}
The McLerran-Venugopalan model describes a highly boosted
hadron/nucleus as a sheet of random color charges which source soft
classical Weizs\"acker-Williams gluon fields. We show that due to
fluctuations, individual configurations are azimuthally
anisotropic. We present initial evidence that impact parameter
dependent small-$x$ JIMWLK resummation preserves such anisotropies
over several units of rapidity. Finally, we compute the first four
azimuthal Fourier amplitudes of the S-matrix of a fundamental dipole
in such background fields.
\end{abstract}

\maketitle

\section{Introduction}
To explain azimuthal asymmetries observed in high-energy pA
collisions~\cite{pPb_ALICE,pPb_ATLAS,pPb_CMS,dAu_RHIC,dAu_STAR}
Refs.~\cite{KovnerLublinsky,Dumitru:2014dra,Dumitru:2014yza} argued
that individual configurations of the light-cone electric fields of
the target should be anisotropic, leading to a non-trivial azimuthal
distribution of a projectile parton scattered off such a target.  That
is, configuration by configuration, two-dimensional rotational
symmetry is broken by ${\bf E}$-field ``domains'' of finite size in
the impact parameter plane.  These, in contrast to Weiss magnetic
domains separated by domain walls, arise purely due to fluctuations of
the valence (large-$x$) random color charge sources for the soft,
small-$x$ ${\bf E}$ field.

Assuming such azimuthal anisotropy of the light-cone electric fields
several features of the data could be described,
at least qualitatively~\cite{Dumitru:2014dra,Dumitru:2014yza}. On the
other hand, a direct calculation of the anisotropic distributions, in
particular for a large nucleus and small $x$ (i.e.\ high energy), has
so far been lacking. It is our goal here to compute scattering of a
dipole off a large nucleus, and specifically, to determine its
angular dependence. That is, we compute the (first four) Fourier
amplitudes of the dipole S-matrix with respect to the azimuthal
orientation of the dipole. We should stress that we do not address the
fluctuations of ${\cal S}(\rt,\bt)$ in impact parameter space $\bt$
(see Ref.~\cite{Schlichting:2014ipa} for a recent study) but rather
its dependence on the size and {\em orientation} of the dipole vector
$\rt$ which is the variable conjugate to the transverse momentum of
the parton in the final state.

\section{The Model}
In the McLerran-Venugopalan model~\cite{MV} the large-$x$ valence
partons are viewed as random, recoilless color charges $\rho^a({\bf
  x})$ which source the semi-classical small-$x$ gluon fields. We
first provide a brief description of how these color charge
configurations are generated on a lattice; more detailed discussions
can be found in the literature~\cite{Krasnitz:1998ns,Lappi:2007ku}.

The effective action describing color charge fluctuations is taken to be
quadratic,
\be \label{eq:S2}
S_{\rm eff}[\rho^a] = \int \ud x^- \ud^2{\bf x} \; 
\frac{\rho^a(x^-,\xt) \, \rho^a(x^-,\xt)}{2\mu^2}
\ee 
with $\mu^2\sim g^2 A^{1/3}$ proportional to the thickness of a
nucleus~\cite{MV}; here $A$ denotes the number of nucleons in the
nucleus. The variance of color charge fluctuations determines the
average saturation scale $Q_s^2 \sim g^4
\mu^2$~\cite{JalilianMarian:1996xn}. The coarse-grained effective
action~(\ref{eq:S2}) applies to (transverse) area elements containing
a large number of large-$x$ ``valence'' charges, $\mu^2 \Delta A_\perp
\sim \Delta A_\perp\,Q_s^2/g^4 \gg 1$.

Hence, in the first step we construct a random configuration of color
charges on a lattice according to the distribution $\exp(-S[\rho])$.
Their (non-Abelian) Weizs\"acker-Williams fields are pure gauges; in
covariant gauge,
\be \label{eq:A+} 
A^{\mu a}(x^-,\xt) = - \delta^{\mu+} \frac{g}{ \nabt^2} \rho^a(x^-,\xt)~.
\ee
This also satisfies $A^-=0$ and thus the only non-vanishing field
strength is $F^{+i}=-\partial^i A^+$. The (light-cone) electric field
is
\be
E^i = \int \ud x^- F^{+i} = - \partial^i \int \ud x^- A^+~.
\ee
The propagation of a fast charge in this field is described by an
eikonal phase given by a light-like SU(3) Wilson line $V(\xt)$:
\be \label{eq:V_rho}
V(\xt) = \mathbb{P} \exp\left\{ ig^2 \int \ud x^-  
\frac{1}{ \nabt^2} t^a \rho^a(x^-,\xt) \right\}, 
\ee
where $\mathbb{P}$ denotes path-ordering in $x^-$. The absolute value
squared of this amplitude gives the S-matrix for scattering of this
charge off the given target field configuration,
\be
{\cal S}_\rho(\rt,\bt) \equiv \frac{1}{N_c}\tr V^\dagger(\xt)\,
V(\yt)~, ~~~~\rt \equiv \xt-\yt~,~~~2\bt \equiv \xt+\yt~.
\ee
Thus, following the ideas leading to the MV model we assume that every
particular scattering event probes one particular configuration in the
target, i.e.\ that the S-matrix is computed with a frozen
$\rho^a(\xt)$. The main purpose of this paper is to analyze the
dependence of the S-matrix on the angular orientation of the dipole
vector $\rt$, conjugate to the transverse momentum, 
 at fixed transverse impact parameter (coordinate) $\bt$.

The S-matrix for a fundamental charge is complex (for three or more
colors). Its real (imaginary) part corresponds to ${\cal C}$-even
(${\cal C}$-odd) exchanges~\cite{CGCodderon}:
\bea
1-D_\rho(\rt)\equiv \mathrm{Re}\,{\cal S}_\rho(\rt) =
 \frac{1}{2N_c}\tr \left[V^\dagger(\xt)\,V(\yt) +
   V^\dagger(\yt)\,V(\xt)\right]~,  \label{eq:Def_D}\\
O_\rho(\rt)\equiv  \mathrm{Im}\,{\cal S}_\rho(\rt) =
 \frac{-i}{2N_c}\tr \left[V^\dagger(\xt)\,V(\yt) -
   V^\dagger(\yt)\,V(\xt)\right]~.  \label{eq:Def_iO}
\eea
${\cal C}$ conjugation transforms $\rho^a(\xt) \to -\rho^a(\xt)$ and
$V(\xt) \to V^*(\xt)$.  The dipole scattering amplitude
$D_\rho(\rt)=D_\rho(-\rt)$ is even under $\rt\to-\rt$ and generates
even azimuthal $v_{2n}$ harmonics while the odderon
$O_\rho(\rt)=-O_\rho(-\rt)$ generates odd
$v_{2n+1}$~\cite{Dumitru:2014dra} of the one-particle distribution.

It is useful to consider the limit of small dipoles, $rQ_s\ll1$. Then
the real part of the S-matrix from Eq.~(\ref{eq:Def_D}) is
\be
\mathrm{Re}\,{\cal S}_\rho(\rt) -1 = \frac{(ig)^2}{2N_c}\tr \left(
\rt\cdot {\bf E}\right)^2 + {\cal O}(r^4)~.
\ee
To compute the elliptic (dipole) asymmetry,
Refs.~\cite{Dumitru:2014dra,KovnerLublinsky,Dumitru:2014yza}
considered the following angular dependence of the two-point function
\be \label{eq:anisoE}
\frac{g^2}{2N_c} \left<\tr E^i(\bt_1) E^j(\bt_2)\right> =
\frac{1}{4} Q_s^2\; \Delta(\bt_1-\bt_2) \, \left(\delta^{ij}
+2{\cal A} \left(\hat{a}^i \hat{a}^j -\frac{1}{2} \delta^{ij}\right) \right)~,
\ee
where $\hat{\bf a}$ corresponds to the ``event plane'' orientation,
and $\Delta(\bt_1-\bt_2)$ describes the ${\bf E}$-field correlations
in the transverse impact parameter plane.
It is implicit that for each configuration ${\bf E}(\bt)$ is rotated
to point in a particular, fixed direction $\hat{\bf a}$ before
performing the ensemble average. In fact, Eq.~(\ref{eq:anisoE}) is
the MV model analogue of the gluon TMD for an unpolarized
target~\cite{TMD,TMD2},
\be
\delta^{ij}f_1^g(x,\kt^2) + 
\left(\hat{k}^i\hat{k}^j-\frac{1}{2}\delta^{ij}\right)
h_1^{\perp g}(x,\kt^2)~.
\ee
Thus, the amplitude ${\cal A}$ from Eq.~(\ref{eq:anisoE}), which we
shall denote $A_2(r)$ below, is basically $h_1^{\perp g}$ at small
$x$. However, beyond the MV model the relation between these functions
may be more involved.

The action~(\ref{eq:S2}) is ${\cal C}$-even and so $\left<
O_\rho(\rt) \right> = 0$ while $\left< D_\rho(\rt) \right>\sim r^2
Q_s^2$ (at small $r$) is proportional to the thickness of the nucleus,
$A^{1/3}$. A ${\cal C}$-odd operator
\be
 \frac{1}{\kappa_3} \, d^{abc} \rho^a \rho^b \rho^c
\ee
with $\kappa_3\sim g^3 A^{2/3}$ could be added to the
action\footnote{Beyond a perturbative treatment of the cubic Casimir one
  would have to add the quartic Casimir, too, so that the action is
  bounded from below~\cite{Petreska}.} which
would then induce an expectation value $\sim A^{1/3}$ for the
odderon~\cite{JV}. This is beyond the scope of the present paper, we
focus here on azimuthal anisotropies due to fluctuations of the charge
densities $\rho^a(\xt)$ and their associated electric fields ${\bf
  E}^a(\xt)$. The main point is that even though the ensemble averaged
S-matrix is isotropic and real (even under charge conjugation) that
fluctuations generate anisotropies and ${\cal C}$-odd contributions
locally in impact parameter space for individual configurations.

%---------------------------------------------------------------------
\section{Implementation}
To generate the random configurations $\rho^a(x^-,\xt)$ via
Monte-Carlo techniques we discretize the longitudinal and transverse
coordinates. The number of sites in the longitudinal direction is
taken to be $N_- = 100$ while the number of sites in either transverse
direction is $N_\perp = 1024$. All of our results presented here have
been obtained with $g^2 \mu a = 0.05$, hence $g^2 \mu L = 51.2$, where
$a\equiv L/N_\perp$ denotes the transverse lattice spacing. We
have determined numerically that $Q_s \approx 0.7125 g^2 \mu$ as
defined from 
\begin{equation}
\left<{\cal S}_\rho\right> (r = \sqrt{2}/Q_s) =
\exp(-1/2).
\label{Qsdef}
\end{equation}
The physical value for the lattice spacing could be
determined by assigning a physical value to $Q_s$; instead, we choose
to measure distance scales in units of $1/Q_s$ or $1/g^2\mu$ and
so this step is not required.

We use periodic boundary conditions in the transverse directions and
solve the Poisson equation~(\ref{eq:A+}) by Fast Fourier
Transform. The amplitude of the zero mode of $\rho^a(\kt)$ is set to
zero before inversion which ensures color neutrality of each
configuration. Alternatively, one could introduce a mass
cutoff $m \ll Q_s$ in the Coulomb propagator in
eq.~(\ref{eq:A+}). Either way, the dynamics of modes with $k$ well
above the IR cutoff is the same.

We have generated about $10^4$ configurations; for each of them we measured
$D_\rho(\rt)$ and $O(\rt)$ at $\bt=0$. Both functions were decomposed
into their Fourier series to extract the amplitudes of azimuthal
anisotropy:
\begin{eqnarray}
&& D_\rho(\rt)  = {\cal N}(r)\;  \left( 1  +  \sum_{n=1}^\infty
  A'_{2n} (r) \cos(2 n \phi_r )  \right)~, \label{eq:Dr_An}\\ 
&&O_\rho(\rt)  = {\cal N}(r)\;  \sum_{n=0}^\infty A'_{2n+1} (r) \cos(
  (2 n+1) \phi_r )~.
\label{Eq:FE}
\end{eqnarray}
The function ${\cal N}(r)$ is the isotropic part of the dipole
S-matrix, see for example Ref.~\cite{Lappi:2007ku}. For a small
dipole, $\langle {\cal N}\rangle(r)\sim \frac{1}{4}r^2 Q_s^2$, up to
logarithms.

The S-matrix ${\cal S}_\rho(\rt)$ rotates randomly from
configuration to configuration. This manifests as a random shift
$\phi_r\to\phi_r-\psi$ in eqs.~(\ref{eq:Dr_An},\ref{Eq:FE}). Hence, on
average over all configurations $\langle A_n'\rangle=0$ for all
$n$. This is a consequence of the fact, already mentioned above,
that the ensemble average of the S-matrix is real and isotropic.

Azimuthal harmonics $v_n$ in hadronic collisions are defined from
multi-particle correlation functions in such a way that they are
invariant under a global shift of the azimuthal angles of all
particles by the same amount. Consequently, we discard this random
phase by defining $A_n = \frac{\pi}{2}|A'_n|$; the factor of $\pi/2$
arises as the inverse average of $\int d\Delta\phi/(2\pi)\, |\cos\,
n\, \Delta\phi| = 2/\pi$. In other
words, we define the amplitudes $A_n$ such that
fluctuations do not average out. We finally obtain their expectation
values over the ensemble of configurations, $\langle A_1 \rangle, \cdots,
\langle A_4 \rangle$, as well as the variances of $A_1$ and $A_2$.

%---------------------------------------------------------------------
\section{Results}
\begin{figure}[htb]
\begin{center}
\includegraphics[height=8cm]{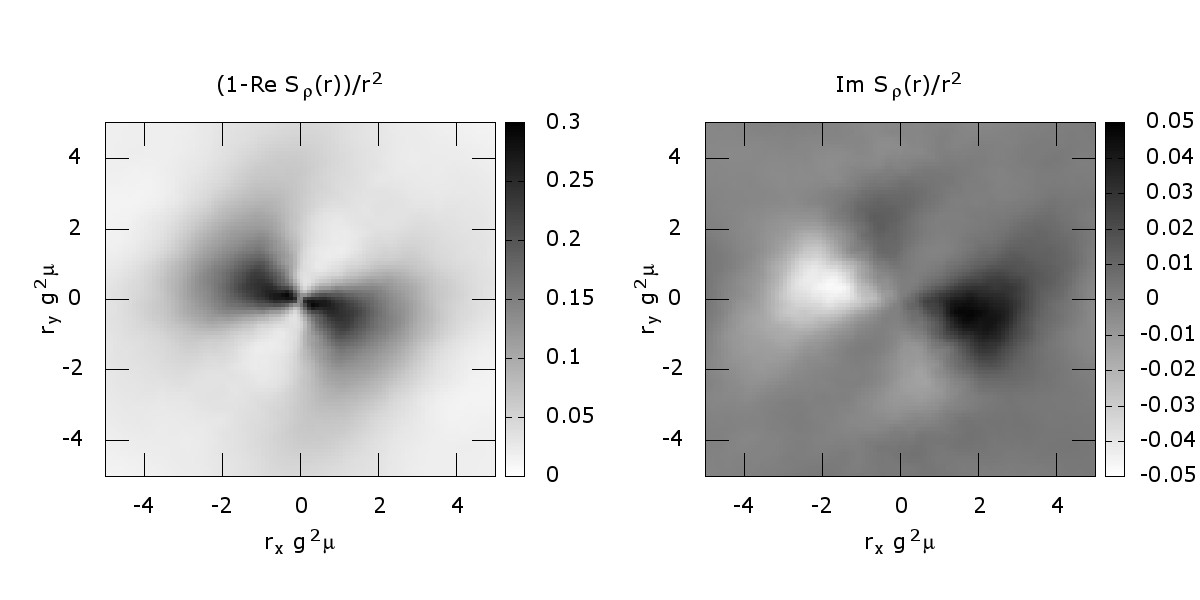}
\end{center}
\vspace*{-1cm}
\caption[a]{The S-matrix in the fundamental representation as a
  function of the dipole vector $\rt = (r_x,r_y)$ at fixed impact
  parameter $\bt=0$ for one particular random configuration of color
  charges $\rho^a(\xt)$.
}
\label{fig:contourConf1}
\end{figure}
\begin{figure}[htb]
\begin{center}
\includegraphics[height=8cm]{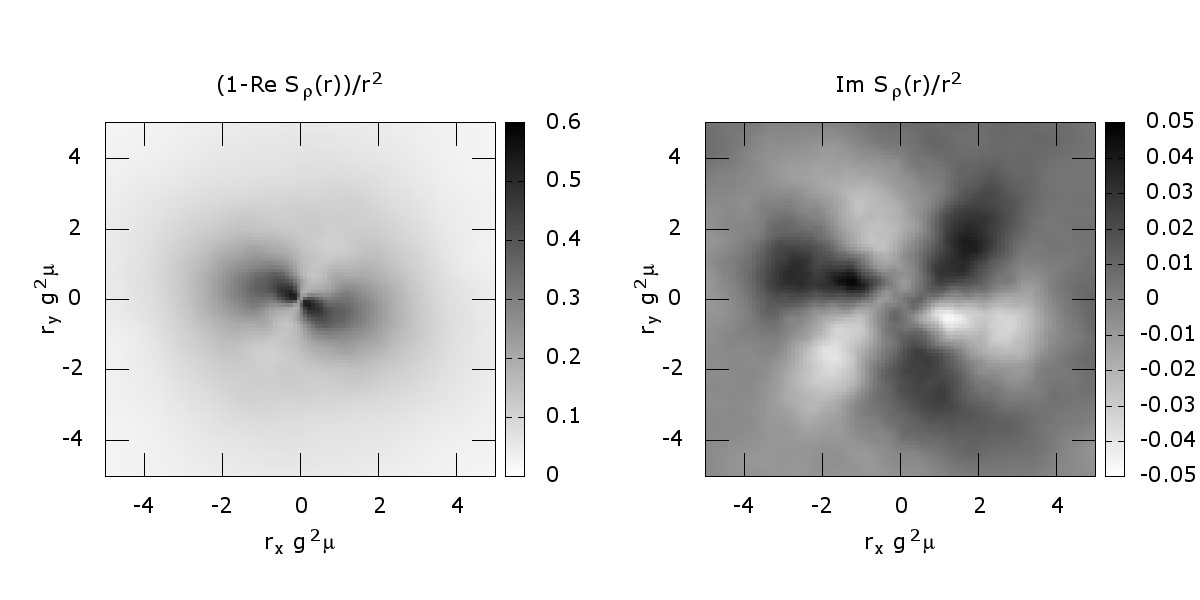}
\end{center}
\vspace*{-1cm}
\caption[a]{Same as Fig.~\ref{fig:contourConf1} for a second
  configuration of color charges $\rho^a(\xt)$.
}
\label{fig:contourConf2}
\end{figure}
Before presenting our results for the azimuthal amplitudes we show two
examples for ${\cal S}_\rho(\rt)$ in Figs.~\ref{fig:contourConf1} and
\ref{fig:contourConf2}. Either of these corresponds to one particular
(random) configuration of color charges. The real parts display
predominantly a $\sim \cos(2\phi)$ angular dependence, with $\phi$ the
angle between $\rt$ and ${\bf E}(\bt=0)$. On the other hand, the
imaginary part for the configuration shown in
Fig.~\ref{fig:contourConf1} is predominantly $\sim \cos(\phi)$ while
that from Fig.~\ref{fig:contourConf2} is mainly $\sim \cos(3\phi)$,
modulo a random phase shift as mentioned above. The figures show,
also, that the angular structures appear at a resolution on the order
of $r\, g^2\mu\sim1$; this is consistent with the requirement $\mu^2\,
\Delta A_\perp\gg1$ mentioned above (which sets the regime of
applicability of the effective theory) at weak coupling: $1/g^2 \gg
1$.

\begin{figure}[htb]
\begin{center}
\includegraphics[height=7cm]{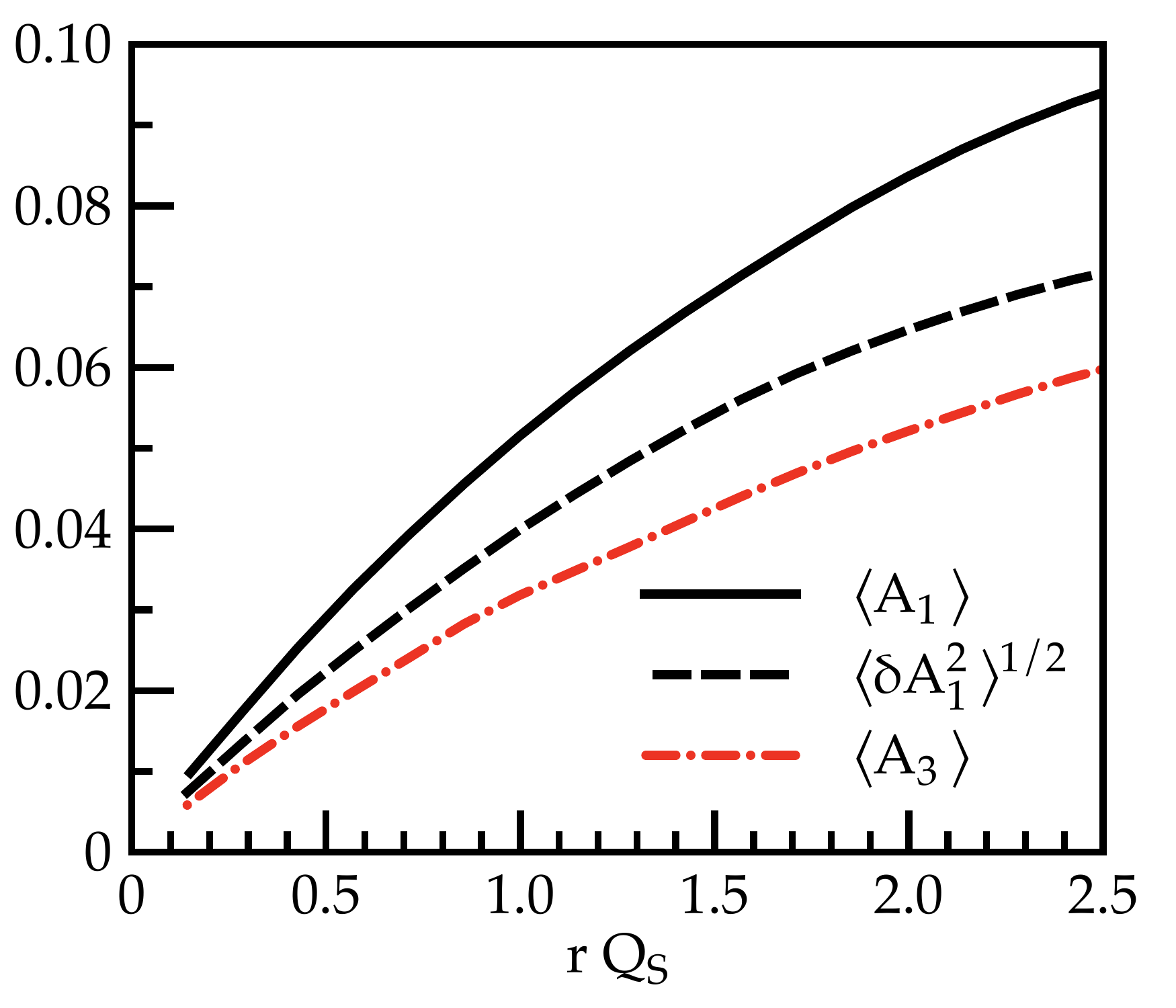}
\includegraphics[height=7cm]{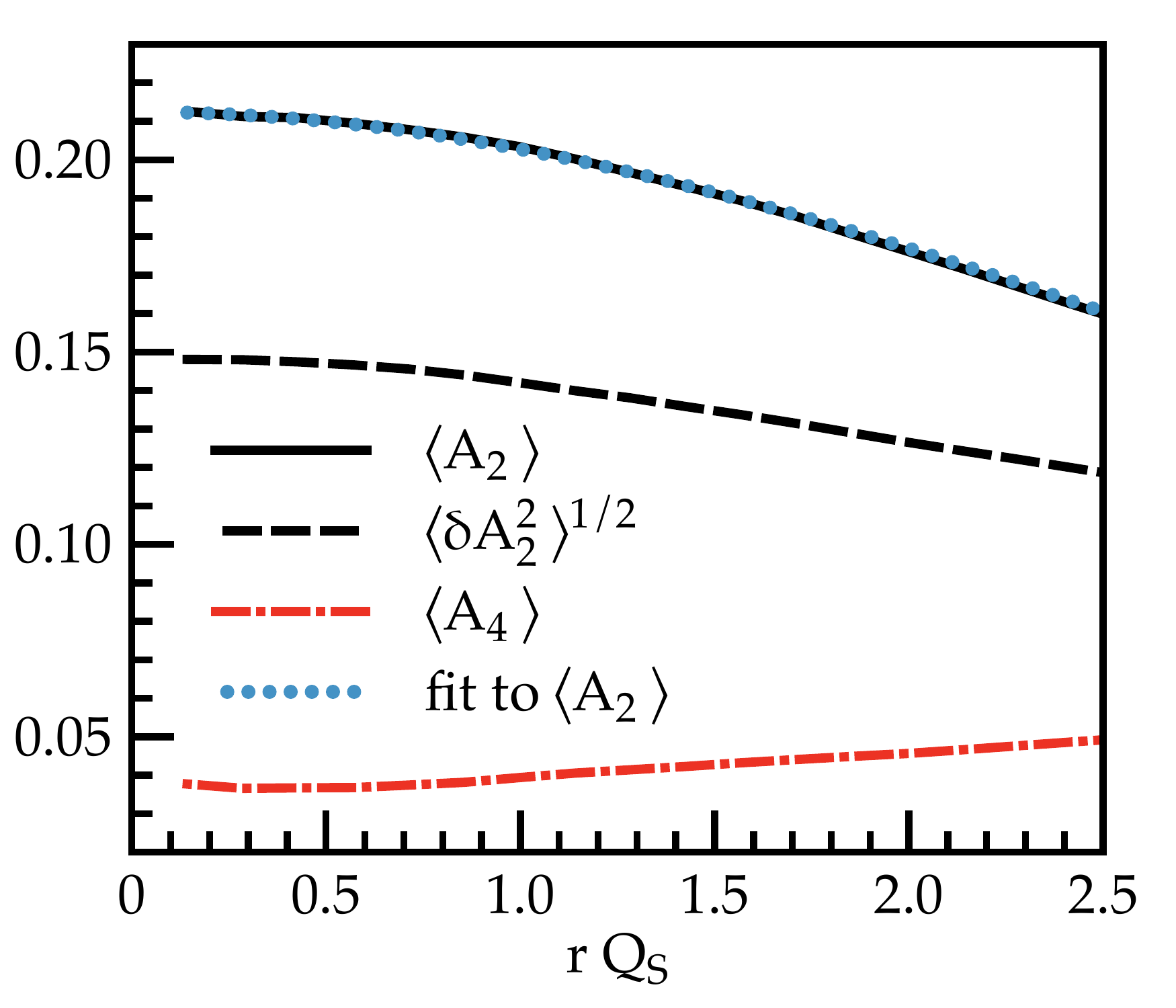}
\end{center}
\vspace*{-0.5cm}
\caption[a]{The averaged amplitudes $\langle A_n\rangle(r)$ vs.\ the
  dipole size $r$ for $n=1,\cdots,4$. The fit to $\langle A_2\rangle$
  corresponds to the function from Eq.~(\ref{Eq:h1perp}).  }
\label{fig:A_n_b0}
\end{figure}
Figure~\ref{fig:A_n_b0} shows our results for the averaged amplitudes of
the first four azimuthal harmonics. As expected, the biggest one is
the quadrupole amplitude $\langle A_2\rangle$ which reaches $\gsim
20\%$ at $r\lsim 1/Q_s$. Such values are in the range of the
asymmetries extracted phenomenologically~\cite{Dumitru:2014dra} for
high-multiplicity p+Pb collisions at LHC energies. However, here we
have not made any attempts to bias the configurations towards ``high
multiplicities''. The fact that the variance $\surd \langle (\delta
A_2)^2\rangle$ is
not much smaller than $\langle A_2\rangle$ indicates that some
configurations generate much larger elliptic asymmetries than
others. Also, we observe that $\langle A_2\rangle$ is approximately
constant for $r<1/Q_s$ since up to quadratic order the real part of
the S-matrix is
\be \label{eq:D_2ndO}
D(\rt) = \frac{g^2}{2N_c} \tr (\rt\cdot{\bf E})^2 
  - \frac{1}{2} \frac{g^4}{4N_c^2} \left[\tr (\rt\cdot{\bf E})^2 \right]^2
  +\cdots
\ee
at small $r$. To derive this expression one performs a gradient
expansion of $\mathrm{Re}\, \tr V(\xt) V^\dagger(\yt)$, assuming that
the electric field is smoothly varying over scales of order $r$.
The leading term on the r.h.s., if scaled by $1/r^2$, is
independent of $r$ which is consistent with the observed constant
$\langle A_2\rangle$ at small $\rt$.

The numerical result for $\langle A_2(r)\rangle$ agrees well with
$h_1^{\perp g}(x, r)$ derived in Ref.~\cite{TMD2}:
\begin{equation}
h_1^{\perp g}(x, \rt^2) \propto \frac{1}{r^2 Q_s^2} \left[ 1 -
  \exp\left( - \frac{r^2 Q_s^2}{4}  \right) \right]~.
\label{Eq:h1perp}
\end{equation}
The agreement suggests that (in the MV model) $\langle
A_2(r)\rangle$ essentially corresponds to the distribution of
linearly polarized gluons, at least for sufficiently small
dipoles. Below we shall see that the functions differ at large $\rt$.

The second term in Eq.~\eqref{eq:D_2ndO} generates a hexadecupole
asymmetry at the next to leading order in $r^2$. However, the
numerical result for $A_4(r)$ shown in Fig.~\ref{fig:A_n_b0} is
essentially constant at small $r$. We interpret this as due to
corrections to the gradient expansion which leads to
Eq.~(\ref{eq:D_2ndO}); a $\sim \cos(4\phi)$ angular component appears
already at ${\cal O}(r^2)$ albeit with a much smaller amplitude than
the $\sim \cos(2\phi)$ harmonic.

We now turn to the odd amplitudes $A_1$ and $A_3$. As already
mentioned above, the expectation value of the odderon over a ${\cal
  C}$-even ensemble such as that generated by the action~(\ref{eq:S2})
is of course zero. Nevertheless, each particular configuration of
semi-classical small-$x$ fields~(\ref{eq:A+}) {\em does} contain a
${\cal C}$-odd component and $iO(\rt)$ as defined in
Eq.~(\ref{eq:Def_iO}) is non-zero. This is due to fluctuations of the
saturation momentum $Q_s$ in impact parameter
space~\cite{Kovchegov:2012ga},
\be
iO(\rt) \sim i\, \alpha_s\, \rt\cdot\nabt_\bt \left(
1 - D(\rt,\bt)\right)
\simeq i\, \alpha_s\, r^3\, Q_s^2 \, Q_c\, {\cal B} \cos \phi_r \left[
1 - \frac{r^2}{4} \left( \frac{Q_c^2\cos^2 \phi_r}{3} + Q_s^2
  \right)
\right]~.
\label{eq:iO_r3}
\ee
The expression on the r.h.s.\ corresponds to an expansion in powers of
$r$; $Q_c$ is a cutoff for the spectrum of fluctuations of $Q_s(\bt)$
which was otherwise assumed to be scale invariant, and ${\cal B}$ is
their amplitude~\cite{Dumitru:2014dra}. Eq.~(\ref{eq:iO_r3}) shows
that for small dipoles, after we divide by the isotropic normalization
factor ${\cal N}(r) \sim r^2$, that we should expect $A_1\sim r$ as
well as a smaller $A_3\sim r^3$. The lattice results appear consistent
with $\langle A_1\rangle\sim r$ at $r\ll1/Q_s$ but so is $\langle
A_3\rangle$, albeit with a smaller slope.  Future simulations on
larger lattices may be able to push to smaller $r$, and the analytical
derivation of Eq.~(\ref{eq:iO_r3}) based on a simple fluctuation
spectrum could perhaps be refined as well.

Just as for the elliptic asymmetry we have also computed the standard
deviation of the amplitude $A_1$. Again, we find that $\surd \langle
(\delta A_1)^2\rangle$ is not much smaller than $\langle A_1\rangle$,
i.e.\ that some configurations generate much larger dipole asymmetries
than others.

We have also analyzed the effect of ``smearing'' the impact parameter
of the projectile over a region corresponding to its size~\cite{smear}. If the
${\bf E}$-field anisotropy exhibits a non-zero correlation length in the
impact parameter
plane~\cite{KovnerLublinsky,Dumitru:2014dra,Dumitru:2014yza},
specifically a correlation length that exceeds the size of the dipole,
then the azimuthal moments should remain approximately  the same.

\begin{figure}[htb]
\begin{center}
\includegraphics[height=7cm]{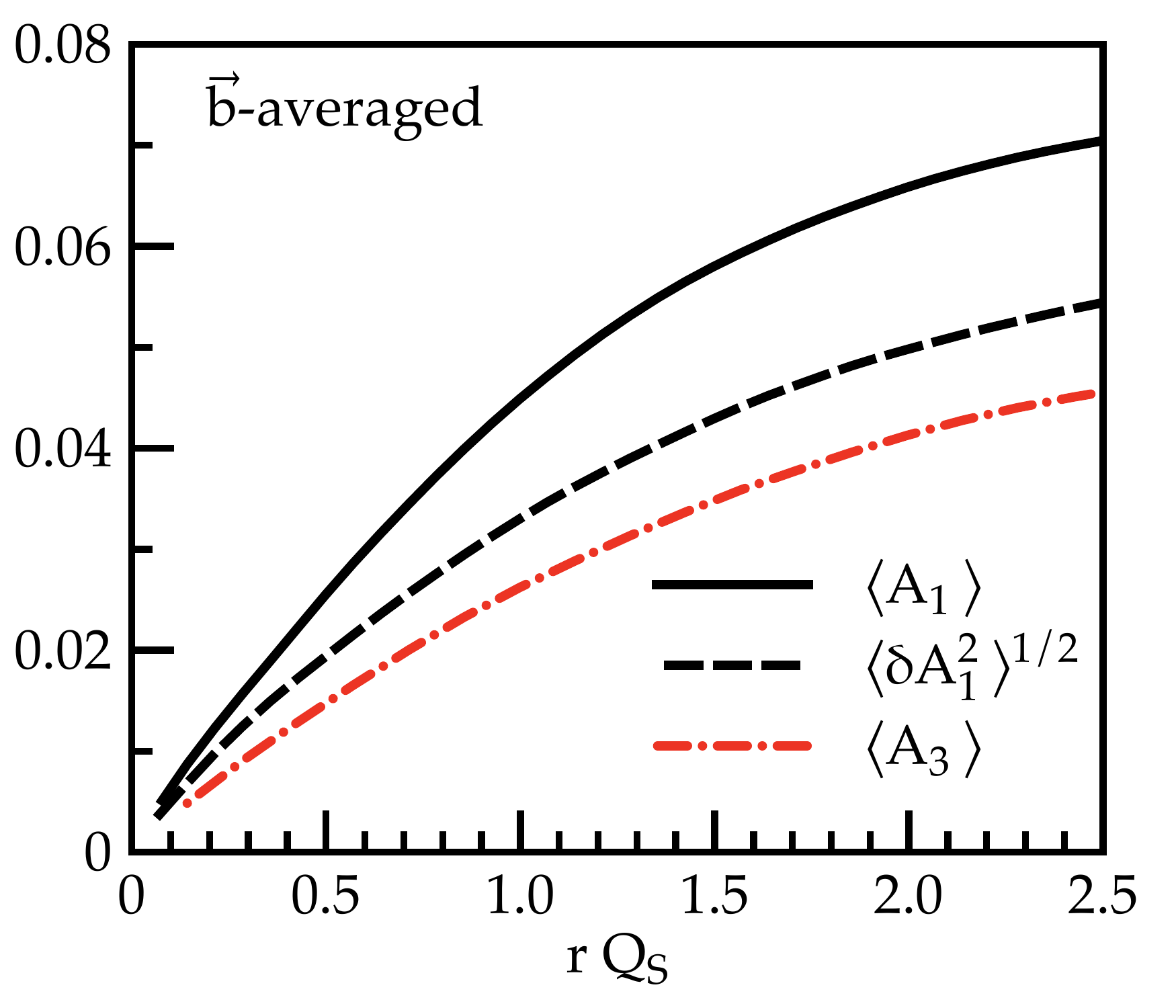}
\includegraphics[height=7cm]{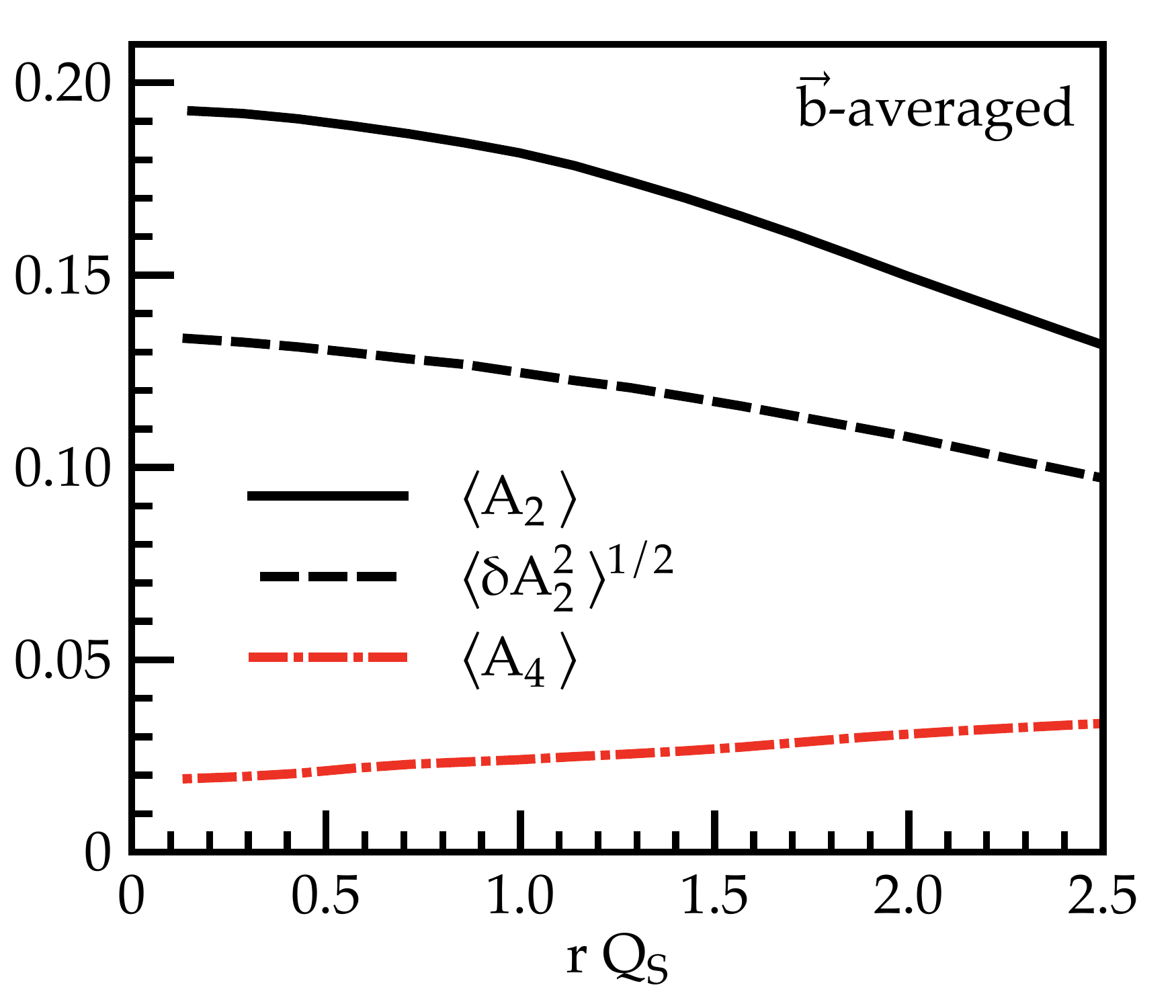}
\end{center}
\vspace*{-0.5cm}
\caption[a]{Same as Fig.~\ref{fig:A_n_b0} (note modified scale on the
  vertical axes) for ``smeared'' S-matrix.
}
\label{fig:A_n_b}
\end{figure}
Hence, we have also computed the azimuthal amplitudes $A_n$ from
``smeared'' configurations:
\be  \label{eq:Dsmear}
\overline D_\rho(\rt,\bt) = \int \frac{\ud^2\bt'}{\pi r^2} \;
\Theta\left(r-|\bt-\bt'|\right)\, D_\rho(\rt,\bt') ~,
\ee
and similarly for $i\overline O_\rho(\rt,\bt)$. On the r.h.s.\ the
points $\xt=\bt'+\rt/2$ and $\yt=\bt'-\rt/2$ are now determined by
$\rt$ and $\bt'$. Equation~(\ref{eq:Dsmear}) averages the
S-matrix over an area $\pi r^2$. The result is shown in
Fig.~\ref{fig:A_n_b} which can be compared to Fig.~\ref{fig:A_n_b0}
from above. Except for a slight suppression of their magnitudes, we do
not observe any substantial modification of the amplitudes $\langle
A_n\rangle$.

\begin{figure}[htb]
\begin{center}
\includegraphics[height=7cm]{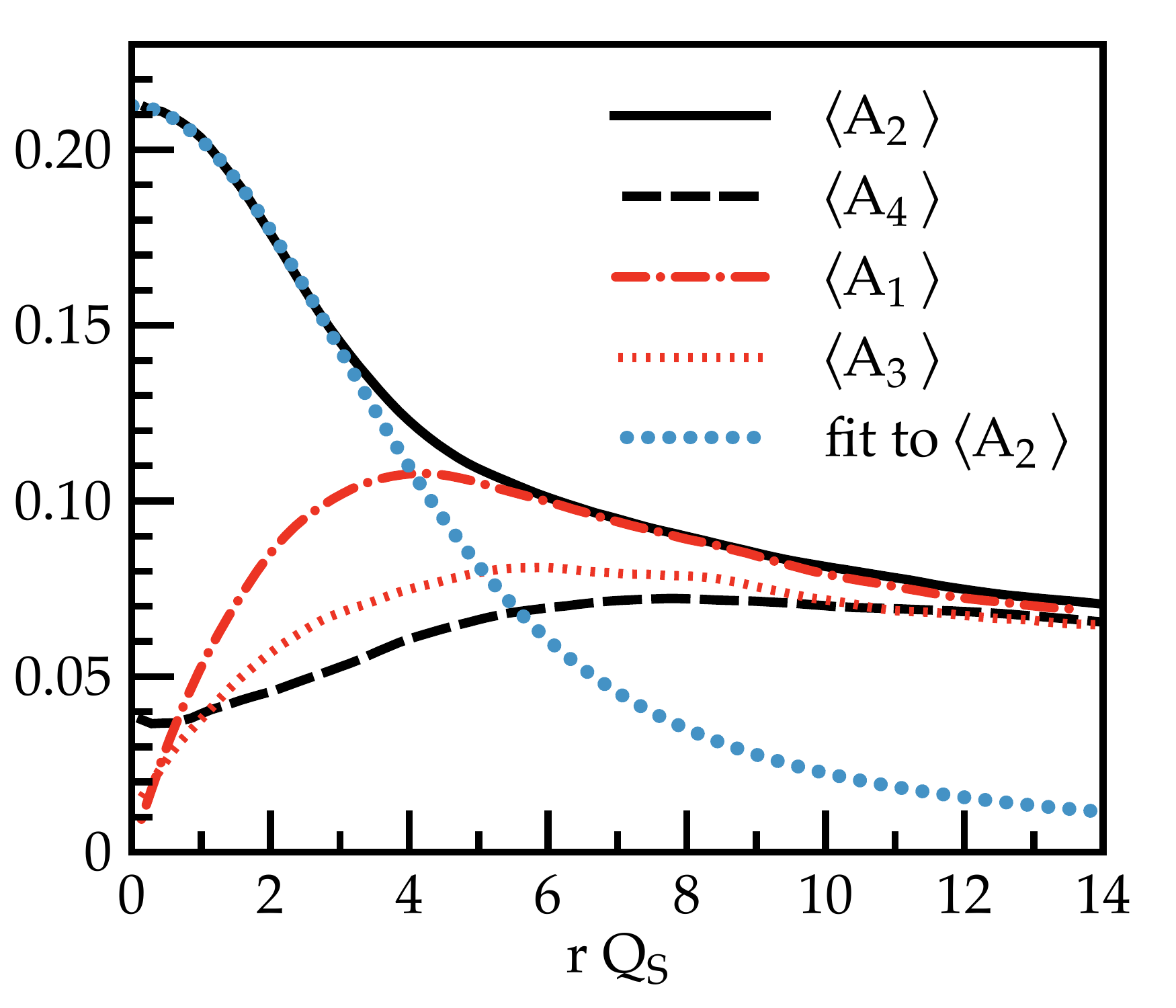}
\includegraphics[height=7cm]{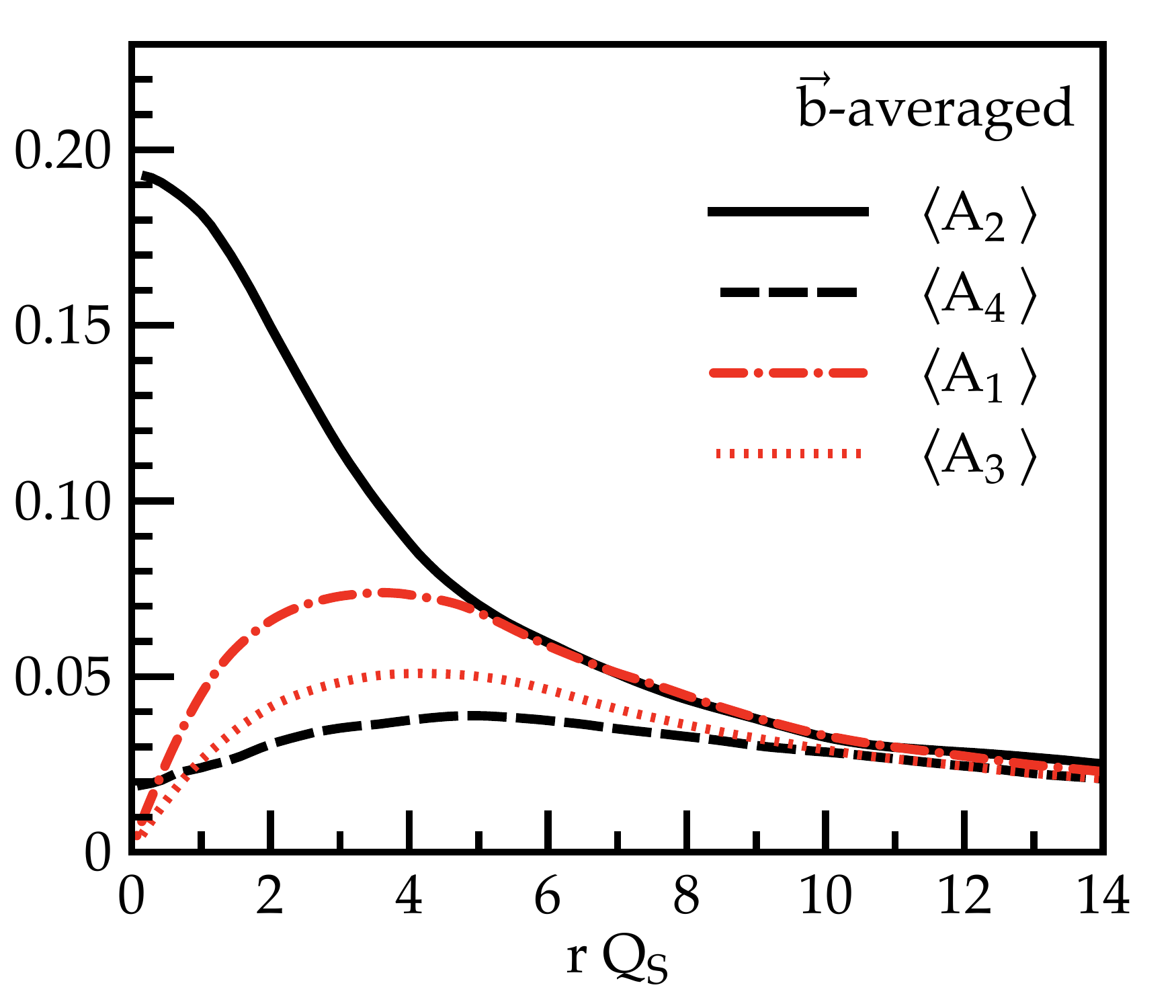}
\end{center}
\vspace*{-0.5cm}
\caption[a]{The averaged amplitudes $\langle A_n\rangle(r)$ vs.\ the
  dipole size $r$ for $n=1,\cdots,4$. This figure focuses on the
  behavior at large $r\gg1/Q_s$. Left: fixed impact parameter
  $\bt=0$; right: impact parameter averaged over an area $\pi r^2$.
}
\label{fig:A_n_largeR}
\end{figure}
The behavior for large dipoles is different,
c.f.\ Fig.~\ref{fig:A_n_largeR}. For a fixed impact parameter the
harmonic amplitudes approach a common non-zero function at large
$r\gg1/Q_s$. This is consistent with universal (angular) scale
invariant fluctuations of the azimuthal dependence of the
S-matrix. Indeed, if $D(\rt,\bt)$ and $O(\rt,\bt)$ are first averaged
over an area $\pi r^2$, see Eq.~(\ref{eq:Dsmear}), then the resulting
$\langle A_n\rangle$ are strongly suppressed. This shows that the
direction of ${\bf E}$ is not correlated over distances far beyond
$\sim 1/Q_s$. Also, as already mentioned above, at large $\rt$ the
function $\langle A_2\rangle(r)$ is seen to differ from $h_1^{\perp g}(x,
\rt^2)$.

We should stress that the behavior of $\left< A_n\right>$ at $r\gg
1/Q_s$ is shown only to reveal their expected universality due to
scale invariant fluctuations (on a circle) within the model used
here. The result applies in the regime far below the lattice IR cutoff
scale $L$ or whichever other IR cutoffs one may choose when
implementing the theory. On the other hand, in practice $Q_s$ is
expected to be on the order of a few GeV only at current collider
energies and so distances of order $10/Q_s$ are not much shorter than
the confinement scale. The MV model used here does not provide a
controlled approximation to QCD near the confinement scale.

\section{High-energy evolution}
In this section we consider effects due to resummation of
boost-invariant quantum fluctuations of the fields. This is done
through the so-called JIMWLK~\cite{jimwlk,JIMWLK_rm1} functional
renormalization group evolution which resums observables to all orders
in $\alpha_s \log(1/x) = \alpha_s Y$.  Performing a step $\Delta Y$ in
rapidity opens phase space for radiation of gluons which modifies the
classical action~(\ref{eq:S2}). This corresponds to a ``random walk'' in
the space of Wilson lines
$V(\xt)$~\cite{JIMWLK_rm1,Blaizot:2002np,Lappi:2012vw}:
\bea
\partial_Y V(\xt) =
V(\xt)
\frac{i}{\pi}\int \ud^2\ut
 \frac{(\xt-\ut)^i\eta^i(\ut)}{(\xt-\ut)^2}
- \frac{i}{\pi}\int \ud^2\vt 
  V(\vt) \frac{(\xt-\vt)^i \eta^i(\vt)}{(\xt-\vt)^2} V^\dag(\vt) 
V(\xt)~.
\label{eq:Lgvn}
\eea
The Gaussian white noise $\eta^i =\eta^i_a t^a$ satisfies
$\langle \eta^a_i(\xt)\rangle =0$ and
\be \label{eq:etaeta}
\langle \eta^a_i(\xt)\; \eta^b_j(\yt)\rangle = \alpha_s \, \delta^{ab}
\delta_{ij}\delta^{(2)}(\xt-\yt).
\ee
The ``left-right symmetric'' form of Eq.~(\ref{eq:Lgvn}) was
introduced in~\cite{Kovner:2005jc,Lappi:2012vw}. We solve the random
walk numerically assuming, for simplicity, a fixed but small coupling
$\alpha_s=0.14$ so that the speed of evolution is not too
rapid\footnote{The ``time'' variable for fixed coupling evolution is
  $\alpha_s Y$.}. Once an ensemble of Wilson lines on the transverse
lattice has been evolved to rapidity $Y$, we can again compute the
dipole scattering amplitude ${\cal S}_Y(\rt)$, its azimuthal Fourier
decomposition and the corresponding saturations scale $Q_s(Y)$ using
Eq.~\eqref{Qsdef}.  We stress that even though we consider a target of
infinite transverse extent, that the evolution equation is solved on a
transverse lattice which does allow for impact parameter dependent
fluctuations.

\begin{figure}[htb]
\begin{center}
\includegraphics[height=7cm]{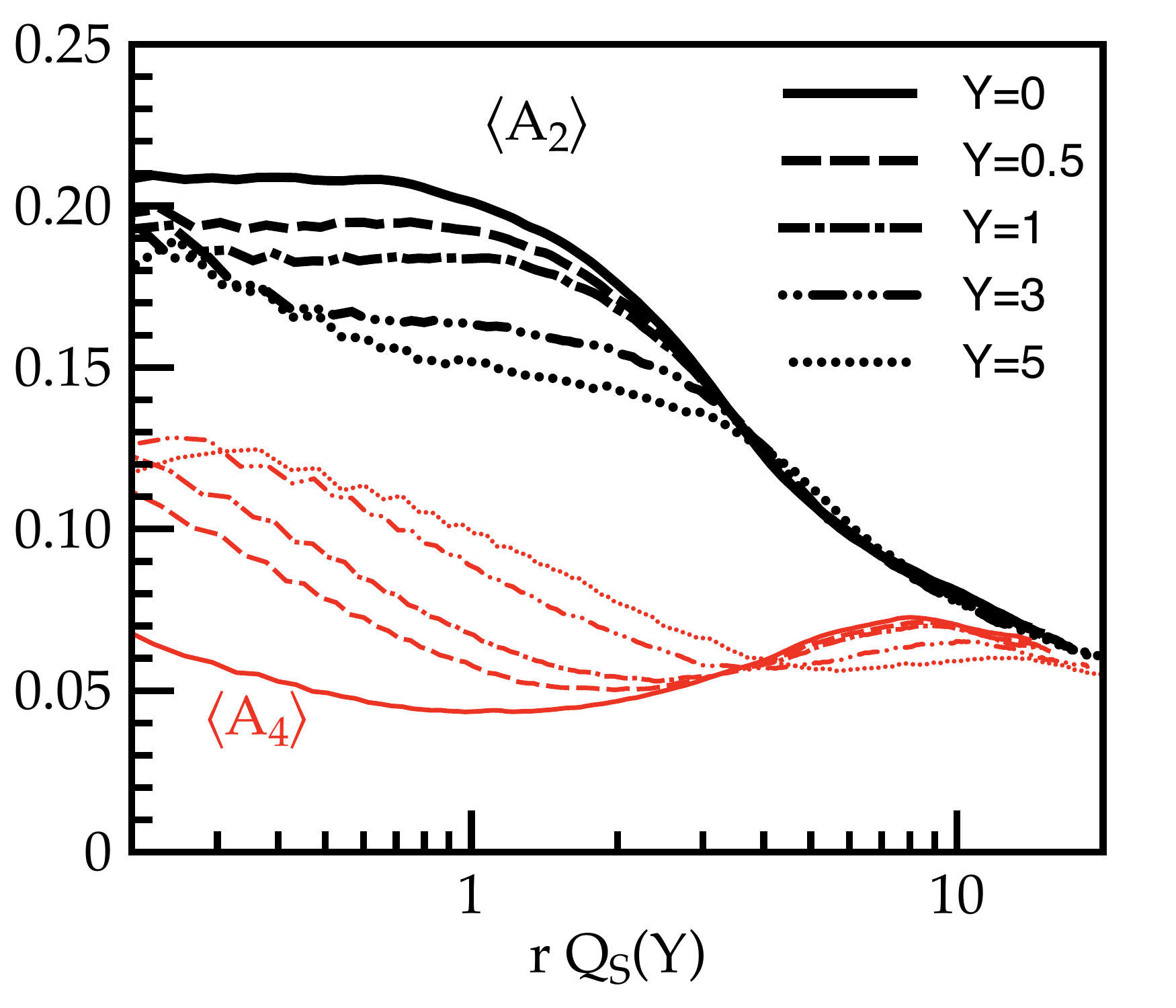}
\includegraphics[height=7cm]{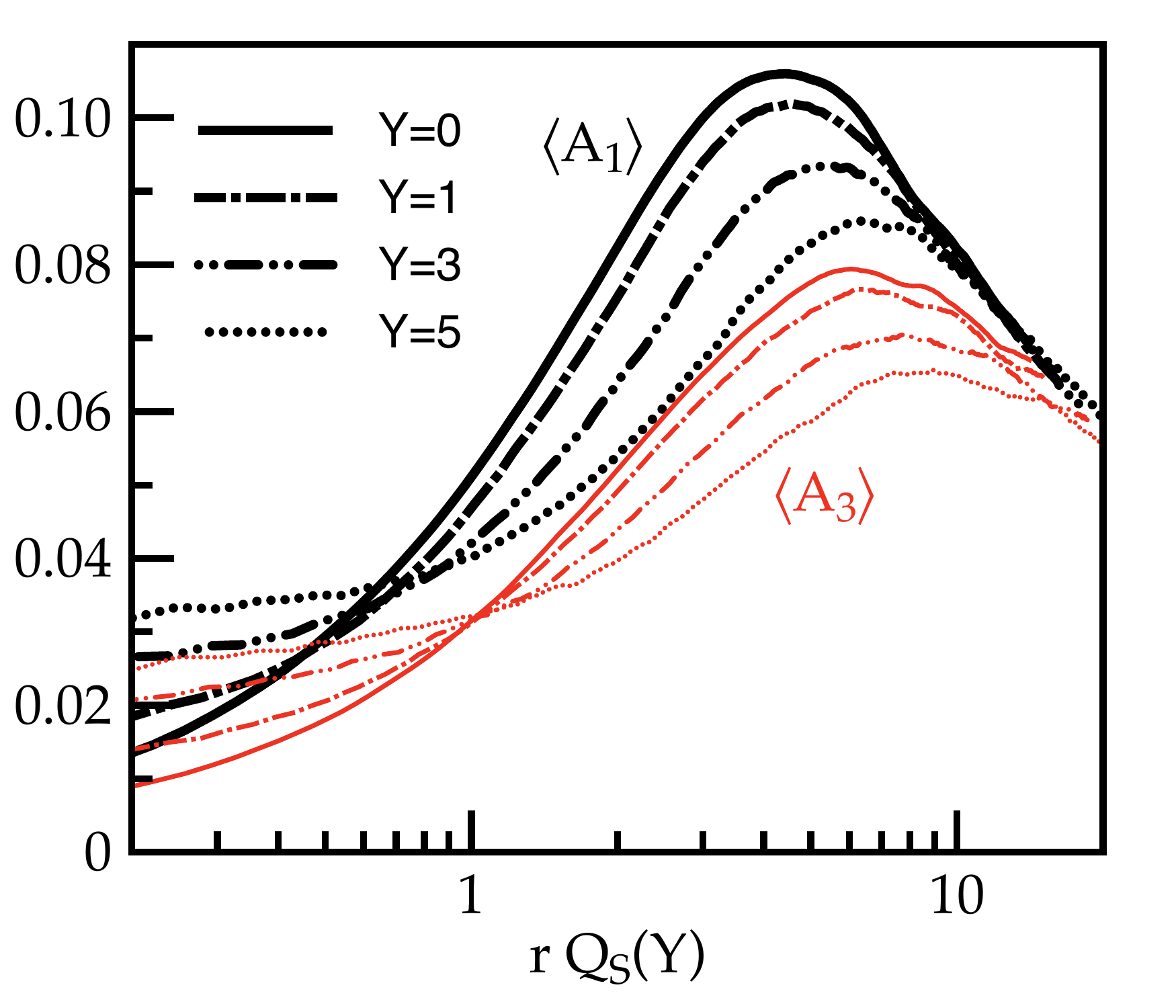}
\end{center}
\vspace*{-0.5cm}
\caption[a]{JIMWLK evolution of $\langle A_2\rangle(r)$ and $\langle
  A_4\rangle(r)$ (left) resp.\ of $\langle A_1\rangle(r)$ and $\langle
  A_3\rangle(r)$ (right). In either plot the lower order harmonic
  corresponds to the upper set of curves.}
\label{fig:A_n_jimwlk}
\end{figure}
In Fig.~\ref{fig:A_n_jimwlk} (left) we show the evolution of $\langle
A_2\rangle(r)$ and $\langle A_4\rangle(r)$. Mean-field evolution of
the dipole has been shown to wash out initial elliptic anisotropies
rather quickly~\cite{KovnerLublinsky}. On the other hand, here we only
observe a relatively slow decrease of $\langle A_2\rangle(r)$ with
$Y$. This is rather intuitive since both the initial anisotropies at
$Y=0$, as well as those of the evolved JIMWLK configurations are
generated by fluctuations of the ``valence charges'' in the transverse
impact parameter plane. Furthermore, we observe that those harmonics
which are small initially, i.e.\ $\langle A_1\rangle(r)$, $\langle
A_3\rangle(r)$ and $\langle A_4\rangle(r)$, in fact increase with
rapidity at small $r$. There is again a universal behavior at very
large $r$.

\section{Summary}
Following the conjecture by Kovner and
Lublinsky~\cite{KovnerLublinsky}, we have analyzed azimuthal
anisotropies of the S-matrix ${\cal S}(\rt)$ for scattering of a
dipole off a large nucleus. They arise due to fluctuations of the
configuration of valence color charges $\rho^a(\xt)$ in the transverse
impact parameter plane\footnote{An alternative picture in terms of
  ${\cal O}(N_c^2)$ fluctuations of the energy-momentum tensor of a
  holographic shock wave has been discussed in
  Ref.~\cite{Noronha:2014vva}. Reference~\cite{Gyulassy:2014cfa}
  considers the azimuthal structure of gluon bremsstrahlung off the
  fast beam-jet sources.}.

For a projectile in the fundamental representation of color SU(3)
these fluctuations generate both charge conjugation even as well as
odd configurations. For small dipoles, $r\lsim 1/Q_s$ the
McLerran-Venugopalan~\cite{MV} model gives $\langle A_2\rangle$ and
$\langle A_4\rangle$ which are approximately constant ($r$
independent). Also, the amplitude of the elliptic harmonic is much
larger than that of the quadrangular harmonic, $\langle A_2\rangle \gg
\langle A_4\rangle$. Odd harmonics appear at higher order in
$r$~\cite{Kovchegov:2012ga,Dumitru:2014dra} and so their amplitudes
decrease with decreasing $r$. The fluctuations of both $A_1$ and $A_2$
are comparable to their mean values, indicating that some
configurations exhibit much larger anisotropies than others.

For large dipoles, $r\gg 1/Q_s$, we find that all amplitudes
$\langle A_1\rangle(r),\cdots,\langle A_4\rangle(r)$ asymptotically
approach a universal function if the S-matrix is evaluated at fixed
impact parameter. This points at angular scale invariant fluctuations
of the direction of ${\bf E}$ over large distances. Accordingly, if
the S-matrix is averaged over an area $\pi r^2$ the resulting
$\cos(n\phi)$ amplitudes are strongly suppressed.

Impact parameter dependent fluctuations during QCD evolution in
rapidity largely preserve the azimuthal amplitudes.  Our calculations
confirm that individual small-$x$ target field configurations do
exhibit angular dependence which would play an important role in
understanding azimuthal $v_n$ harmonics in pp and pA
collisions~\cite{KovnerLublinsky,Dumitru:2014dra,Dumitru:2014yza}. In
particular, the amplitude of elliptic anisotropies $\langle
A_2\rangle\sim 15-20\%$ is on the order of the $v_2$ harmonic observed
in p+Pb collisions at the LHC. Moreover, similar studies as the one
performed here might be able to shed some light on the behavior of the
linearly polarized gluon distribution $h_1^{\perp g}(x, \rt^2)$ at
small $x$; work in that direction is in progress.

\begin{acknowledgments}
We thank B.~Schenke, S.~Schlichting and L.~McLerran for useful discussions.
A.D.\ gratefully acknowledges support by the DOE Office
of Nuclear Physics through Grant No.\ DE-FG02-09ER41620 and from The
City University of New York through the PSC-CUNY Research Award
Program, grant 67119-0045.
V.S.\ thanks D.\ Shubina for discussions and for providing a C++ library 
for color matrices. The numerical computations were performed at 
the High Performance Computing Center, Michigan State University.
\end{acknowledgments}

%---------------------------------------------------------------------

\end{document}